\def\ls{\lower4pt\hbox{${\buildrel < \over \sim}$}}
\def\gs{\lower4pt\hbox{${\buildrel > \over \sim}$}}

\documentclass[12pt, preprint]{aastex}

\slugcomment{Submitted to {\it The Astrophysical Journal}}

\shorttitle{The WEBT campaign on 3C~279 in 2006}
\shortauthors{B\"ottcher et al.}

\begin{document}

\title{The WEBT Campaign on the Blazar 3C~279 in 2006\footnote{For 
questions regarding the availability of the data from the WEBT campaign 
presented in this paper, please contact the WEBT President Massimo 
Villata at {\tt villata@oato.inaf.it}}}

\author{M. B\"ottcher\altaffilmark{2}, S. Basu\altaffilmark{2}, 
M. Joshi\altaffilmark{2}, M. Villata\altaffilmark{3}, 
A. Arai\altaffilmark{4},
N. Aryan\altaffilmark{5},
I. M. Asfandiyarov\altaffilmark{6},
U. Bach\altaffilmark{3},
R. Bachev\altaffilmark{7},
A. Berduygin\altaffilmark{8},
M. Blaek\altaffilmark{9},
C. Buemi\altaffilmark{12},
A. J. Castro-Tirado\altaffilmark{11},
A. De Ugarte Postigo\altaffilmark{11},
A. Frasca\altaffilmark{12},
L. Fuhrmann\altaffilmark{30,13,3},
V. A. Hagen-Thorn\altaffilmark{17},
G. Henson\altaffilmark{14},
T. Hovatta\altaffilmark{16},
R. Hudec\altaffilmark{9},
M. Ibrahimov\altaffilmark{6},
Y. Ishii\altaffilmark{4},
R. Ivanidze\altaffilmark{15},
M. Jel\'inek\altaffilmark{11},
M. Kamada\altaffilmark{4},
B. Kapanadze\altaffilmark{15}
M. Katsuura\altaffilmark{4},
D. Kotaka\altaffilmark{4},
Y. Y. Kovalev\altaffilmark{30,31},
Yu. A. Kovalev\altaffilmark{31},
P. Kub\'anek\altaffilmark{9},
M. Kurosaki\altaffilmark{4},
O. Kurtanidze\altaffilmark{15},
A. L\"ahteenm\"aki\altaffilmark{16},
L. Lanteri\altaffilmark{3},
V. M. Larionov\altaffilmark{17},
L. Larionova\altaffilmark{17}
C.-U. Lee\altaffilmark{18},
P. Leto\altaffilmark{10},
E. Lindfors\altaffilmark{8},
E. Marilli\altaffilmark{12},
K. Marshall\altaffilmark{19},
H. R. Miller\altaffilmark{19},
M. G. Mingaliev\altaffilmark{32},
N. Mirabal\altaffilmark{20},
S. Mizoguchi\altaffilmark{4},
K. Nakamura\altaffilmark{4},
E. Nieppola\altaffilmark{16},
M. Nikolashvili\altaffilmark{15},
K. Nilsson\altaffilmark{8},
S. Nishiyama\altaffilmark{4},
J. Ohlert\altaffilmark{21},
M. A. Osterman\altaffilmark{19},
S. Pak\altaffilmark{23},
M. Pasanen\altaffilmark{8},
C. S. Peters\altaffilmark{24},
T. Pursimo\altaffilmark{25},
C. M. Raiteri\altaffilmark{3}, 
J. Robertson\altaffilmark{26},
T. Robertson\altaffilmark{27}
W. T. Ryle\altaffilmark{19},
K. Sadakane\altaffilmark{4},
A. Sadun\altaffilmark{5},
L. Sigua\altaffilmark{15}
B.-W. Sohn\altaffilmark{18},
A. Strigachev\altaffilmark{7},
N. Sumitomo\altaffilmark{4},
L. O. Takalo\altaffilmark{8},
Y. Tamesue\altaffilmark{4},
K. Tanaka\altaffilmark{4},
J. R. Thorstensen\altaffilmark{24},
G. Tosti\altaffilmark{13},
C. Trigilio\altaffilmark{12},
G. Umana\altaffilmark{12},
S. Vennes\altaffilmark{26},
S. Vitek\altaffilmark{11},
A. Volvach\altaffilmark{28},
J. Webb\altaffilmark{29}
M. Yamanaka\altaffilmark{4},
H.-S. Yim\altaffilmark{17},
}

\altaffiltext{2}{Astrophysical Institute, Department of Physics and Astronomy, \\
Clippinger 339, Ohio University, Athens, OH 45701, USA}
\altaffiltext{3}{Istituto Nazionale di Astrofisica (INAF), Osservatorio Astronomico di Torino,\\
Via Osservatorio 20, I-10025 Pino Torinese, Italy}
\altaffiltext{4}{Astronomical Institute, Osaka Kyoiku University, Kashiwara-shi, \\
Osaka, 582-8582 Japan}
\altaffiltext{5}{Department of Physics, University of Colorado at Denver, \\
Campus Box 157, P. O. Box 173364, Denver, CO 80217-3364, USA}
\altaffiltext{6}{Ulugh Beg Astronomical Institute, Academy of Sciences of Uzbekistan, \\
33 Astronomical Str., Tashkent 700052, Uzbekistan}
\altaffiltext{7}{Institute of Astronomy, Bulgarian Academy of Sciences,\\
72 Tsarigradsko Shosse Blvd., 1784 Sofia, Bulgaria}
\altaffiltext{8}{Tuorla Observatory, University of Turku, 21500 Piikki\"o, Finland}
\altaffiltext{9}{Astronomical Institute, Academy of Sciences of the Czech Republic,\\
CZ-251 65 Ondrejov, Czech Republic}
\altaffiltext{10}{Istituto di Radioastronomia, Sezione di Noto, C. da Renna Bassa -- \\
Loc. Casa di Mezzo C. P. 141, I-96017 Noto, Italy}
\altaffiltext{11}{Instituto de Astrofisica de Andalucia, Apartado de Correos, 3004, \
E-18080 Granada, Spain}
\altaffiltext{12}{Osservatorio Astrofisico di Catania, Viale A.\ Doria 6, \\
I-95125 Catania, Italy}
\altaffiltext{13}{Osservatorio Astronomico, Universit\`a di Perugia, Via B.\ Bonfigli, \\
I-06126 Perugia, Italy}
\altaffiltext{14}{East Tennessee State University and SARA Observatory, \\
Department of Physics, Astronomy, and Geology, Box 70652, Johnson City, TN 37614}
\altaffiltext{15}{Abastumani Observatory, 383762 Abastumani, Georgia}
\altaffiltext{16}{Mets\"ahovi Radio Observatory, Helsinki University of Technology, \\
Mets\"ahovintie 114, 02540 Kylm\"al\"a, Finland}
\altaffiltext{17}{Astronomical Institute, St. Petersburg State University, \\
Universitetsky pr.\ 28, Petrodvoretz, 198504 St. Petersburg, Russia}
\altaffiltext{18}{Korea Astronomy \& Space Science Institute, 61-1 Whaam-Dong, Yuseong-Gu,\\
Daejeon 305-348, Korea}
\altaffiltext{19}{Department of Physics and Astronomy, Georgia State University,\\
Atlanta, GA 30303, USA}
\altaffiltext{20}{Department of Astronomy, University of Michigan, \\
830 Dennison Building, Ann Arbor, MI 48109-1090, USA}
\altaffiltext{21}{Michael Adrian Observatory, Astronomie-Stiftung Trebur, \\
Fichtenstraße 7, D-65468 Trebur, Germany}
\altaffiltext{23}{Department of Astronomy and Space Science, Kyung Hee University, \\
Seocheon, Gilheung, Yongin, Gyeonggi, 446-701, South Korea}
\altaffiltext{24}{Department of Physics and Astronomy, Dartmouth College, MS 6127, \\
Hannover, NH 03755, USA}
\altaffiltext{25}{Nordic Optical Telescope, Apartado 474, E-38700 Santa Cruz de La Palma,\\
Santa Cruz de Tenerife, Spain}
\altaffiltext{26}{Florida Institute of Technology and SARA Observatory, 150 West University Boulevard, \\
Melbourne, FL 32901-6975, USA}
\altaffiltext{27}{Ball State University and SARA Observatory, Department of Physics and Astronomy, \\
Munice, IN 47306, USA}
\altaffiltext{28}{Crimean Astrophysical Observatory, Nauchny, Crimea 98409, Ukraine}
\altaffiltext{29}{Florida International University and SARA Observatory,\\
University Park Campus, Miami, FL 33199, USA}
\altaffiltext{30}{Max-Planck-Institut f\"ur Radioastronomie, Auf dem H\"ugel 69,\\
D-53121 Bonn, Germany}
\altaffiltext{31}{Astro Space Center of Lebedev Physical Institute, Profsoyuznaya 84/32,\\
Moscow 117997, Russia}
\altaffiltext{32}{Special Astrophysical Observatory, Nizhnij Arkhyz, Karachai-Cherkessia 369167, Russia}

\begin{abstract}
The quasar 3C~279 was the target of an extensive multiwavelength 
monitoring campaign from January through April 2006. An optical-IR-radio 
monitoring campaign by the Whole Earth Blazar Telescope (WEBT) 
collaboration was organized around Target of Opportunity X-ray
and soft $\gamma$-ray observations with {\it Chandra} and {\it INTEGRAL}
in mid-January 2006, with additional X-ray coverage by {\it RXTE} and 
{\it Swift} XRT. In this paper we focus on the results of the WEBT 
campaign. 

The source exhibited substantial variability of optical flux and spectral 
shape, with a characteristic time scale of a few days. The variability 
patterns throughout the optical BVRI bands were very closely correlated 
with each other, while there was no obvious correlation between the 
optical and radio variability. After the ToO trigger, the optical flux 
underwent a remarkably clean quasi-exponential decay by about one magnitude, 
with a decay time scale of $\tau_d \sim 12.8$~d. 

In intriguing contrast to other (in particular, BL~Lac type) blazars, we
find a lag of shorter-wavelength behind longer-wavelength variability throughout 
the RVB wavelength ranges, with a time delay increasing with increasing frequency. 
Spectral hardening during flares appears delayed with respect to a rising 
optical flux. This, in combination with the very steep IR-optical continuum 
spectral index of $\alpha_o \sim 1.5$ -- 2.0, may indicate a highly oblique 
magnetic field configuration near the base of the jet, leading to inefficient 
particle acceleration and a very steep electron injection spectrum. 

An alternative explanation through a slow (time scale of several days)
acceleration mechanism would require an unusually low magnetic field of 
$B \lesssim 0.2$~G, about an order of magnitude lower than inferred from 
previous analyses of simultaneous SEDs of 3C~279 and other FSRQs with 
similar properties. 

\end{abstract}


\section{Introduction}

Flat-spectrum radio quasars (FSRQs) and BL~Lac objects are 
active galactic nuclei (AGNs) commonly unified in the class 
of blazars. They exhibit some of the most violent high-energy
phenomena observed in AGNs to date. Their spectral energy
distributions (SEDs) are characterized by non-thermal continuum 
spectra with a broad low-frequency component in the radio -- UV 
or X-ray frequency range and a high-frequency component from
X-rays to $\gamma$-rays. Their electromagnetic radiation exhibits 
a high degree of linear polarization in the optical and radio 
bands and rapid variability at all wavelengths. Radio interferometric 
observations often reveal radio jets with individual components 
exhibiting apparent superluminal motion. At least episodically, 
a significant portion of the bolometric flux is emitted in 
$> 100$~MeV $\gamma$-rays. 46 blazars have been detected 
and identified with high confidence in high energy ($> 100$~MeV) 
$\gamma$-rays by the {\it Energetic Gamma-Ray Experiment Telescope
(EGRET)} instrument on board the {\it Compton Gamma-Ray Observatory} 
\citep[CGRO,][]{hartman99,mhr01}. 

In the framework of relativistic jet models, the low-frequency (radio
-- optical/UV) emission from blazars is interpreted as synchrotron
emission from nonthermal electrons in a relativistic jet. The
high-frequency (X-ray -- $\gamma$-ray) emission could either be
produced via Compton upscattering of low frequency radiation by the
same electrons responsible for the synchrotron emission \citep[leptonic
jet models; for a recent review see, e.g.,][]{boettcher07a}, or 
due to hadronic processes initiated by relativistic protons 
co-accelerated with the electrons \citep[hadronic models, for 
a recent discussion see, e.g.,][]{muecke01,muecke03}. 

The quasar 3C279 ($z = 0.538$) is one of the best-observed flat 
spectrum radio quasars, not at last because of its prominent 
$\gamma$-ray flare shortly after the launch of {\it CGRO} 
in 1991. It has been persistently detected by {\it EGRET} 
each time it was observed, even in its very low quiescent 
states, e.g., in the winter of 1992 -- 1993, and is 
known to vary in $\gamma$-ray flux by roughly two orders of 
magnitude \citep{maraschi94,wehrle98}. It has been monitored
intensively at radio, optical, and more recently also X-ray
frequencies, and has been the subject of intensive multiwavelength
campaigns \citep[e.g.,][]{maraschi94,hartman96,wehrle98}. 

Also at optical wavelengths, 3C~279 has exhibited substantial 
variability over up to two orders of magnitude ($R \sim 
12.5$ -- 17.5). Variability has been observed on a 
variety of different time scales, from years, down to 
intra-day time scales. The most extreme variability patterns
include intraday variability with flux decays of $\lesssim 
0.1^{\rm mag}$/hr \citep{kb07}. Observations with the {\it 
International Ultraviolet Explorer} in the very low activity 
state of the source in December 1992 -- January 1993 revealed 
the existence of a thermal emission component, possibly related 
to an accretion disk, with a luminosity of $L_{\rm UV} \sim 
2 \times 10^{46}$~erg~s$^{-1}$ if this component is assumed 
to be emitting isotropically \citep{pian99}. \cite{pian99} have 
also identified an X-ray spectral variability trend in archival 
{\it ROSAT} data, indicating a lag of $\sim 2$ -- 3~days of the 
soft X-ray spectral hardening behind a flux increase. Weak evidence
for spectral variability was also found within the {\it EGRET}
(MeV -- GeV) energy range \citep{nandi07}. At low $\gamma$-ray flux
levels, an increasing flux seems to be accompanied by a spectral
softening, while at high flux levels, no consistent trend was
apparent.

The quasar 3C~279 was the first object in which superluminal motion was
discovered \citep{whitney71,cotton79,unwin89}. Characteristic apparent
speeds of individual radio components range up to $\beta_{\rm app} 
\sim 17$ \citep{cotton79,homan03,jorstad04}, indicating pattern flow 
speeds with bulk Lorentz factors of up to $\Gamma \sim 17$. Radio jet 
components have occasionally been observed not to follow straight, 
ballistic trajectories, but to undergo slight changes in direction
between parsec- and kiloparsec-scales \citep{homan03,jorstad04}. 
VLBA polarimetry indicates that the electric field vector is 
generally well aligned with the jet direction on pc to kpc scales 
\citep{jorstad04,ojha04,lh05,helmboldt07}, indicating that the magnetic field 
might be predominantly perpendicular to the jet on those length scales. 

A complete compilation and modeling of all available SEDs simultaneous 
with the 11 {\it EGRET} observing epochs has been presented in \cite{hartman01a}. 
The modeling was done using the time-dependent leptonic (SSC + EIC)
model of \cite{bms97,bb00} and yielded quite satisfactory fits for
all epochs. The results were consistent with other model fitting works 
\citep[e.g.,][]{bednarek98,sikora01,moderski03} concluding that the 
X-ray -- soft $\gamma$-ray portion of the SED might be dominated by 
SSC emission, while the {\it EGRET} emission might require an additional, 
most likely external-Compton, component. The resulting best-fit 
parameters were consistent with an increasing bulk Lorentz factor,
but decreasing Lorentz factors of the ultrarelativistic electron 
distribution in the co-moving frame of the emission region during 
$\gamma$-ray high states, as compared to lower $\gamma$-ray states 
\citep{hartman01a}. However, such an interpretation also required
changes of the overall density of electrons, and the spectral index
of the injected electron power-law distribution, which did not show
any consistent trend with $\gamma$-ray luminosity. 

\cite{hartman01b} have investigated cross correlations between 
different wavelength ranges, in particular, between optical,
X-ray, and $\gamma$-ray variability. In that work, a general
picture of a positive correlation between optical, X-ray and 
$\gamma$-ray activity emerged, but no consistent trends of time 
lags between the different wavelength ranges were found.

The discussion above illustrates that, in spite of the intensive 
past observational efforts, the physics driving the broadband
spectral variability properties of 3C~279 are still rather poorly 
understood. For this reason, \cite{collmar07b} proposed an
intensive multiwavelength campaign in an optical high state of
3C~279, in order to investigate its correlated radio -- IR
-- optical -- X-ray -- soft $\gamma$-ray variability. The campaign
was triggered on Jan. 5, 2006, when the source exceeded an R-band 
flux corresponding to R = 14.5. It involved intensive radio, near-IR
(JHK), and optical monitoring by the Whole Earth Blazar Telescope
\cite[WEBT\footnote{\tt http://www.to.astro.it/blazars/webt}, see, 
e.g.][and references therein]{raiteri06,villata07} collaboration 
through April of 2006, focusing on a core period of Jan. and Feb. 
2006. In order to illustrate the source's behaviour leading up to 
the trigger in January 2006, previously unpublished radio and optical 
data from late 2005 are also included in the analysis presented in this 
paper. X-ray and soft $\gamma$ observations were carried out by all 
instruments on board the {\it International Gamma-Ray Astrophysics
Laboratory (INTEGRAL)} during the period of Jan. 13 -- 20, 2006.
Additional, simultaneous X-ray coverage was obtained by {\it 
Chandra} and {\it Swift} XRT. These observations were supplemented 
by extended X-ray monitoring with the {\it Rossi X-Ray Timing 
Explorer (RXTE)}. In this paper, we present details of the 
data collection, analysis, and results of the WEBT (radio -- IR 
-- optical) campaign. Preliminary results of the multiwavelength 
campaign have been presented in \cite{collmar07a} and 
\cite{boettcher07a,boettcher07b}, and a final, comprehensive 
report on the result of the entire multiwavelength campaign 
will appear in \cite{collmar07b}.

Throughout this paper, we refer to $\alpha$ as the energy 
spectral index, $F_{\nu}$~[Jy]~$\propto \nu^{-\alpha}$. A 
cosmology with $\Omega_m = 0.3$, $\Omega_{\Lambda} = 0.7$, 
and $H_0 = 70$~km~s$^{-1}$~Mpc$^{-1}$ is used. In this cosmology,
and using the redshift of $z = 0.538$, the luminosity distance 
of 3C~279 is $d_L = 3.1$~Gpc. 

\begin{figure}
\plotone{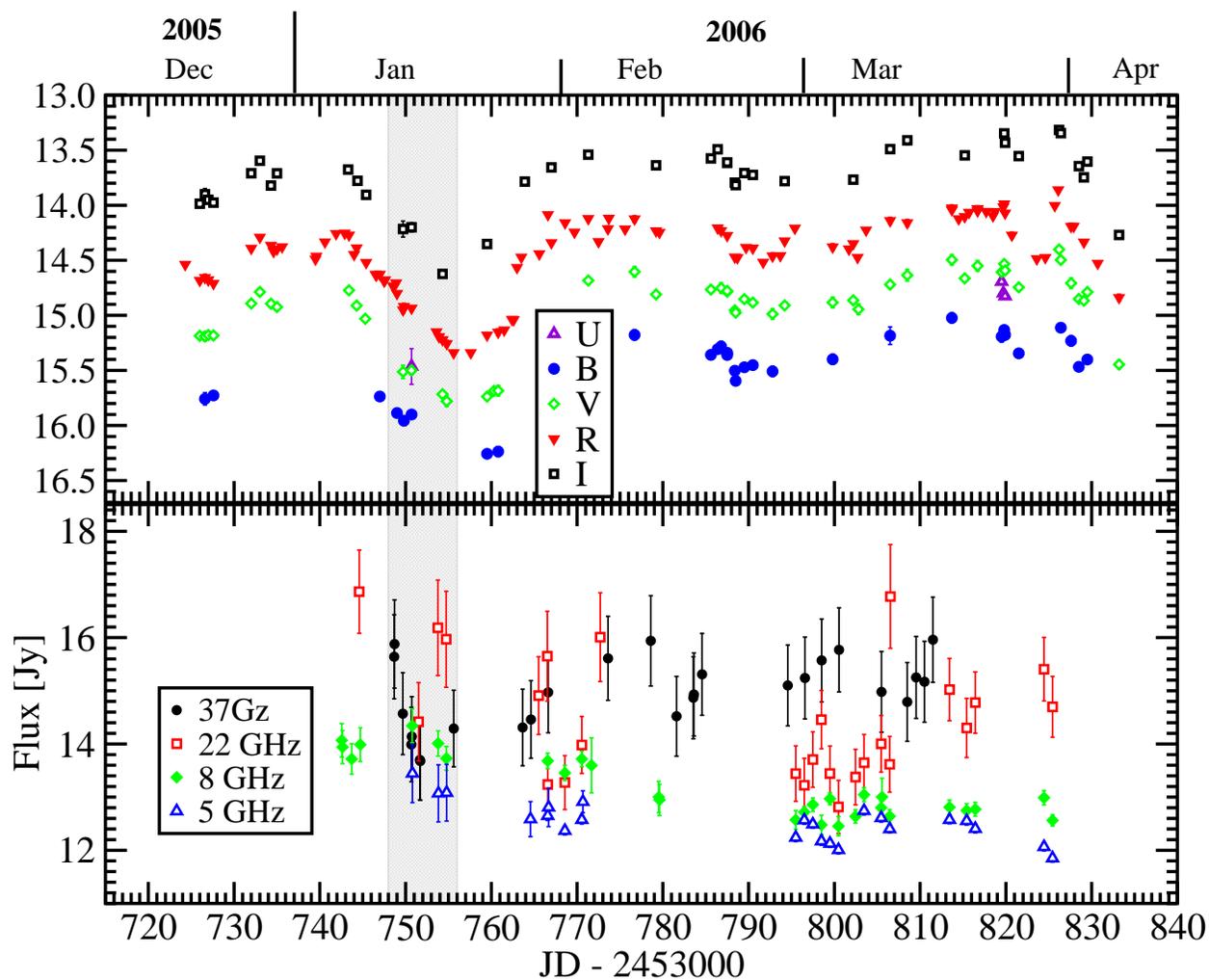}
\caption{Timeline of the broadband campaign on 3C~279 in 2006,
including the optical and radio light curves during the entire
campaign period. The gray shaded area indicates the period of the 
{\it INTEGRAL, Chandra}, and {\it Swift} observations.}
\label{timeline}
\end{figure}

\section{\label{observations}Observations, data reduction, 
and light curves}

3C~279 was observed in a coordinated multiwavelength campaign 
at radio, near-IR, optical (by the WEBT collaboration), 
X-ray ({\it Chandra}, {\it Swift}, {\it RXTE PCA}, {\it INTEGRAL 
JEM-X}), and soft $\gamma$-ray ({\it INTEGRAL}) energies. The overall 
timeline of the campaign, along with the measured long-term light 
curves at radio and optical frequencies is illustrated in Fig. 
\ref{timeline}. Simultaneous X-ray coverage with all X-ray / soft
$\gamma$-ray telescopes mentioned above was obtained in the time
frame Jan. 13 -- 20, as indicated by the gray shaded area in Fig. 
\ref{timeline}. Detailed results of those high-energy observations 
will be presented in \cite{collmar07b}. Table \ref{observatories} 
lists all participating observatories which contributed data to 
the WEBT campaign. In total, 25 ground-based radio, infrared, 
and optical telescopes in 12 countries on 4 continents contributed 
2173 data points. 

\begin{deluxetable}{lccc}
\tabletypesize{\scriptsize}
\tablecaption{List of observatories that contributed data to the 
WEBT campaign}
\tablewidth{0pt}
\tablehead{
\colhead{Observatory} & \colhead{Specifications} & \colhead{frequency / filters} &
\colhead{$N_{\rm obs}$}
}
\startdata
\multispan 4 \hss \bf Radio \hss \\
\noalign{\smallskip\hrule\smallskip}
Mets\"ahovi, Finland                & 14 m  & 37 GHz           & 70 \\
Medicina, Italy                     & 32 m  & 5, 8, 22 GHz     & 32 \\
Noto, Italy                         & 32 m  & 8, 22 GHz        & 6  \\
RATAN-600, Russia                   & 576 m (ring) & 1, 2.3, 5, 8, 11, 22 GHz & 138 \\
Crimean Astr. Obs., Ukraine (RT-22) & 22 m  & 36 GHz           & 7  \\
\noalign{\smallskip\hrule\smallskip}
\multispan 4 \hss \bf Infrared \hss \\
\noalign{\smallskip\hrule\smallskip}
Roque (NOT), Canary Islands & 2.56 m & J, H, K & 3 \\
\noalign{\smallskip\hrule\smallskip}
\multispan 4 \hss \bf Optical \hss \\
\noalign{\smallskip\hrule\smallskip}
Abastumani, Georgia (FSU)     & 70 cm         & R             & 127 \\
ARIES, Naintal, India         &               & R             & 63 \\
Belogradchik, Bulgaria        & 60 cm         & V, R, I       & 75 \\
BOOTES-1, Spain               & 30 cm         & R             & 151 \\
Catania, Italy                & 91 cm         & U, B, V       & 33 \\
Crimean Astr. Obs., Ukraine   & 70 cm         & B, V, R, I    & 47 \\
SMARTS, CTIO, Chile           & 90 cm         & B, V, R       & 33 \\
Kitt Peak (MDM), Arizona, USA & 130 cm        & U, B, V, R, I  & 190 \\
Kitt Peak (MDM), Arizona, USA & 240 cm        & R             & 77 \\
Michael Adrian Obs., Germany  & 120 cm        & R             & 9 \\
Mt. Lemmon, Arizona, USA      & 100 cm        & B, V, R, I    & 214 \\
Mt. Maidanak (AZT-22), Uzbekistan & 150 cm        & B, V, R, I    & 44 \\
Osaka Kyoiku, Japan           & 51 cm         & V, R, I       & 494 \\
Roque (KVA), Canary Islands   & 35 cm         & R             & 75 \\
Roque (NOT), Canary Islands   & 256 cm        & U, B, V, R, I & 7 \\
SARA, Arizona, USA             & 90 cm         & B, V, R, I    & 242 \\
Tenagra, Arizona, USA         & 81 cm         & B, V, R, I    & 19 \\
Torino, Italy                 & 105 cm        & B, V, R       & 3 \\
Tuorla, Finland               & 103 cm        & R             & 84 \\
\noalign{\smallskip\hrule}
\enddata
\label{observatories}
\end{deluxetable}

\subsection{\label{optical}Optical and near-infrared observations}

The observing strategy and data analysis followed to a large extent 
the standard procedure for the optical data reduction for WEBT campaigns
which is briefly outlined below. For more information on standard data 
reduction procedures for WEBT campaigns see also: 
\cite{villata00,raiteri01,villata02,boettcher03,villata04a,villata04b,raiteri05,boettcher05}

It had been suggested that, optimally, observers perform photometric 
observations alternately in the B and R bands, and include complete 
(U)BVRI sequences at the beginning and the end of each observing
run. Exposure times should be chosen to obtain an optimal compromise
between high precision (instrumental errors less than $\sim 0.03$~mag
for small telescopes and $\sim 0.01$~mag for larger ones) and high time
resolution. If this precision requirement leads to gaps of 15 -- 20
minutes in each light curve, we suggested to carry out observations 
in the R band only. Observers were asked to perform bias and dark
corrections as well as flat-fielding on their frames, and obtain 
instrumental magnitudes, applying either aperture photometry (using 
IRAF or CCDPHOT) or Gaussian fitting for the source 3C~279 and four
recommended comparison stars. This 
calibration has then been used to convert instrumental to standard 
photometric magnitudes for each data set. In the next step, unreliable 
data points (with large error bars at times when higher-quality data 
points were available) were discarded. Our data did not provide evidence 
for significant variability on sub-hour time scales. Consequently, error 
bars on individual data sets could be further reduced by re-binning 
on time scales of typically 15 -- 20~min. 
The data resulting at this stage of the analysis are displayed in
the top panel of Fig. \ref{timeline}.

In order to provide information on the intrinsic broadband 
spectral shape (and, in particular, a reliable extraction 
of B - R color indices), the data were then de-reddened using 
the Galactic Extinction coefficients of \cite{schlegel98}, 
based on $A_B = 0.123$~mag and $E(B-V) = 
0.029$~mag\footnote{\tt http://nedwww.ipac.caltech.edu/}.

Possible contaminations of the optical color information could
generally also arise from contributions from the host galaxy
and the optical -- UV emission from an accretion disk around
the central supermassive black hole in 3C~279. However, these
contributions are not expected to be significant in the case
of our campaign data: Assuming absolute magnitudes of $M_V 
\sim -23$ and $M_B \sim -21$ for typical quasar host galaxies
at $z \sim 0.5$ \citep[e.g.][]{floyd04,zakamska06}, their 
contribution in the V and B band, respectively, at the distance 
of 3C~279 would be $V_{\rm gal} \sim 19.5$ and $B_{\rm gal} 
\sim 21.6$, respectively. These are at least about four magnitudes
fainter than the actually measured total B and V magnitudes 
during our campaign, and thus negligible. The possible 
contribution of an accretion disk can be estimated on the 
basis of the thermal component for which \cite{pian99} found 
evidence in IUE observations during the 1992/1993 low state 
of 3C~279. Their best fit to this component suggests $U \sim 
18.6$ and $B \sim 21$ (and much fainter contributions at lower 
frequencies), which corresponds to a contribution of $\lesssim 
2.5$~\% to the total B and U magnitudes measured during our 
campaign. Therefore, both the host galaxy and the accretion 
disk contribution are neglected in our further analysis.

The only infrared observations obtained for this campaign were
one sequence of JHK exposures taken on January 15, 2006, with 
the 2.56~m NOT on Roque de los Muchachos on the Canary Island 
of La Palma. The resulting fluxes are included in the SED 
displayed in Fig. \ref{SED}.

\subsubsection{\label{opt_lightcurves}Optical light curves}

The optical (and radio) light curves from December 2005 to April 
2006 are displayed in Fig. \ref{timeline}. The densest coverage 
was obtained in the R band, and the figure clearly indicates 
that the variability in B, V, and I bands closely tracks the 
R-band behaviour. The coverage in the U band was extremely sparse
and does not allow any assessment of the U-band light curve during 
our campaign. Therefore, the U-band will be ignored in the following 
discussion, and we will describe the main features of the variability
behavior based on the R-band light curve. 

The optical light curves show variability with magnitude changes of 
typically $\lesssim 0.5^{\rm mag}$ on time scales of a few days. The
most notable exception to this relatively moderate variability is the
major dip of the brightnesses in all optical bands right around our 
coordinated X-ray / soft $\gamma$-ray observations in January 2006 
($\approx$~JD~2453742 -- 2453770). In the R-band, the light curve followed
an unusually clean exponential decay over 1.1 mags. in 13 days, i.e.,
a slope of $dR/dt = 0.085$~mag/day or a flux decay as $F(t) = F(t_0) 
\, e^{-(t - t_0) / \tau_d}$ with a decay constant of $\tau_d = 12.8$~d. 
Only moderate intraday deviations on a characteristic scale of 
$\lesssim 0.1$~mag/d are superposed on this smooth exponential 
decay. 

In contrast to the smooth decline of the optical brightness during 
January 6 -- 20, 2006, the subsequent re-brightening to levels 
comparable to those before the dip, appears much more erratic 
and involves a remarkably fast rise by 
$\sim 0.5^{\rm mag}$ within $\sim 1$~d on Jan. 27 (JD 2453763). 
Unfortunately, the detailed shape of this fast rise was not well 
sampled in our data set. If this was a quasi-exponential rise with 
a slope of $dR/dt \sim 0.5$~mag/d, it would correspond to a rise 
time scale of $\tau_r \sim 2.2$~d.

\subsection{\label{radio}Radio observations}

At radio frequencies, 3C~279 was monitored using the 14~m Mets\"ahovi 
Radio Telescope of the Helsinki University of Technology, at 37~GHz, 
the 32-m radio telescope of the Medicina Radio Observatory near Bologna, 
Italy, at 5, 8, and 22~GHz, the 32-m antenna of the Noto Radio Observatory
on Sicily, Italy, at 8 and 22~GHz, the 576-m ring telescope (RATAN-600) 
of the Russian Academy of Sciences, at 1, 2.3, 5, 8, 11, and 22~GHz,
and the 22-m RT-22 dish at the Crimean Astrophysical Observatory,
Ukraine, at 36~GHz. 

The Mets\"ahovi data have been reduced with the standard procedure 
described in \cite{ter98}. The resulting 37~GHz light curve is reasonably 
well sampled during the period mid-January -- mid-March 2006. Inspection by
eye in comparison to the optical light curves displayed in Fig. \ref{timeline}
appears to indicate that the optical and 37~GHz light curves follow similar
variability patterns with a radio lead before the optical variability 
by $\sim 5$~days. However, a discrete cross-correlation analysis between 
the R-band and 37~GHz radio light curves did not reveal a significant
signal to confirm this suggestion. 

Also included in Fig. \ref{timeline} are the radio light curves at 5, 8, 
and 22~GHz. For details of the analysis of data from the Medicina and 
Noto radio observatories at those frequencies, see \cite{bach07}. As 
already apparent in Fig. \ref{timeline}, most of the data at frequencies 
below 37~GHz were not well sampled on the $\lesssim 3$-months time scale 
of the 2006 campaign, and any evidence for variability did not show a 
discernable correlation with the variability at higher (radio and optical) 
frequencies. 

All radio data contributed to this campaign are stored in the 
WEBT archive (see {\tt http://www.to.astro.it/blazars/webt/}
for information regarding availability of the data).
Radio data at all observed frequencies have been included in the 
quasi-simultaneous SED for Jan 15, 2006, shown in Fig. \ref{SED}. 
Given the generally very moderate radio variability at frequencies 
below 37~GHz, a linear interpolation between the two available data 
points nearest in time to Jan. 15, 2006, was used to construct an 
estimate of the actual radio fluxes at that time.

\begin{figure}
\plotone{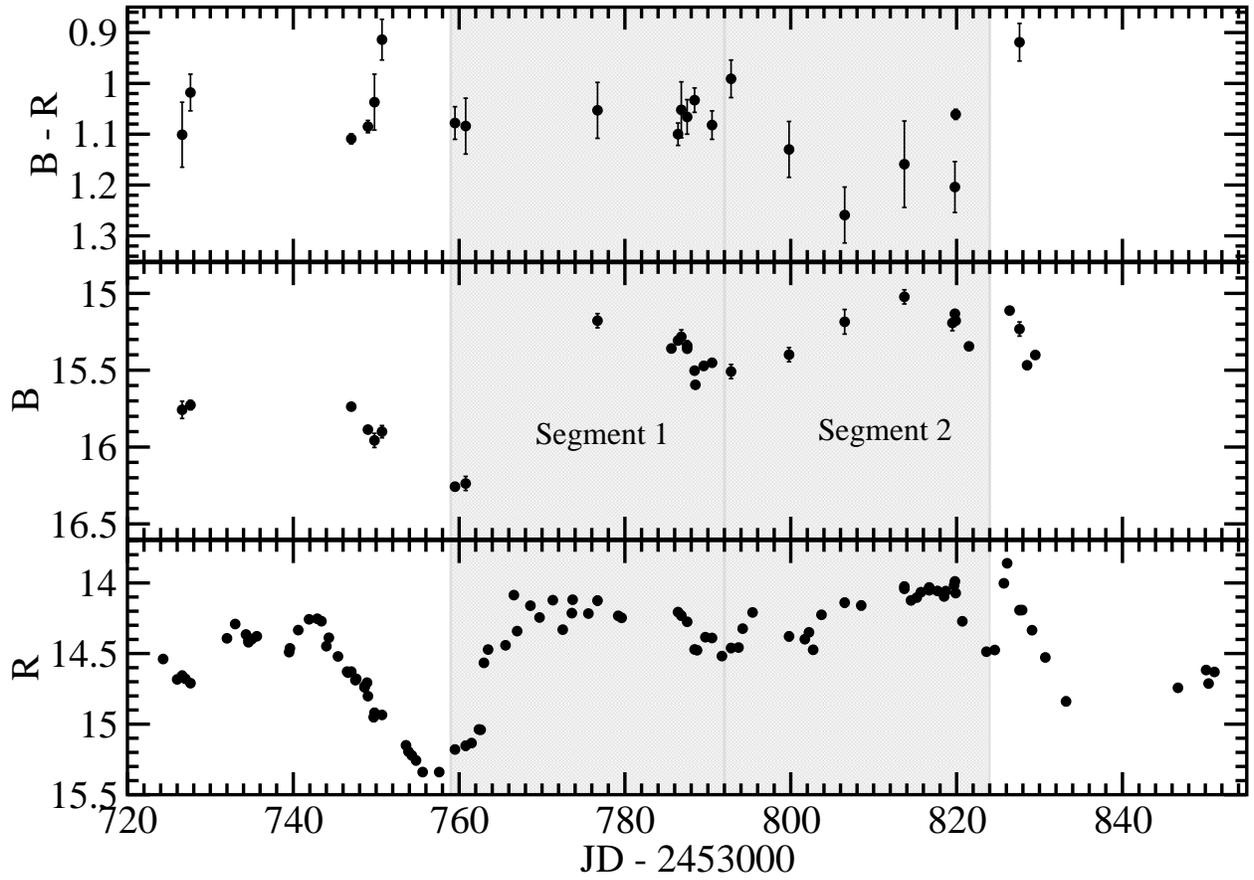}
\caption{Light curve of the B and R magnitudes and the B -- R color index 
of 3C~279 over the duration of the entire campaign. }
\label{B_R_lc}
\end{figure}

\section{\label{variability}Optical spectral variability}

In this section, we will describe spectral variability
phenomena, i.e. the variability of spectral (and color) 
indices and their correlations with monochromatic source fluxes. 
We will concentrate here on the optical spectral variability as
indicated by a change of the optical color. In particular, our
observing strategy was optimized to obtain a good sampling of
the B -- R color index as a function of time. Since our data 
did generally not indicate substantial flux changes on sub-hour
time scales, we extracted B -- R color indices wherever both 
magnitudes were available within 20 minutes of each other. 
Fig. \ref{B_R_lc} shows the B -- R color history over the entire
campaign, compared to the B and R band light curves. Overall,
there is no obvious correlation between the light curves and
the color behavior of the source on long time scales. However,
two short-term sequences attracted our attention: There is a
sequence of brightness decline accompanied by a spectral hardening 
(declining B -- R index) around JD 2453750 (Jan. 14, 2006), and 
another sequence of a brightness increase accompanied by a 
spectral softening around JD 2453790 -- 2453806 (Feb. 23 -- 
Mar. 11). However, we caution that incomplete sampling, in
particular in the B-band may introduce spurious effects. 

\begin{figure}
\plotone{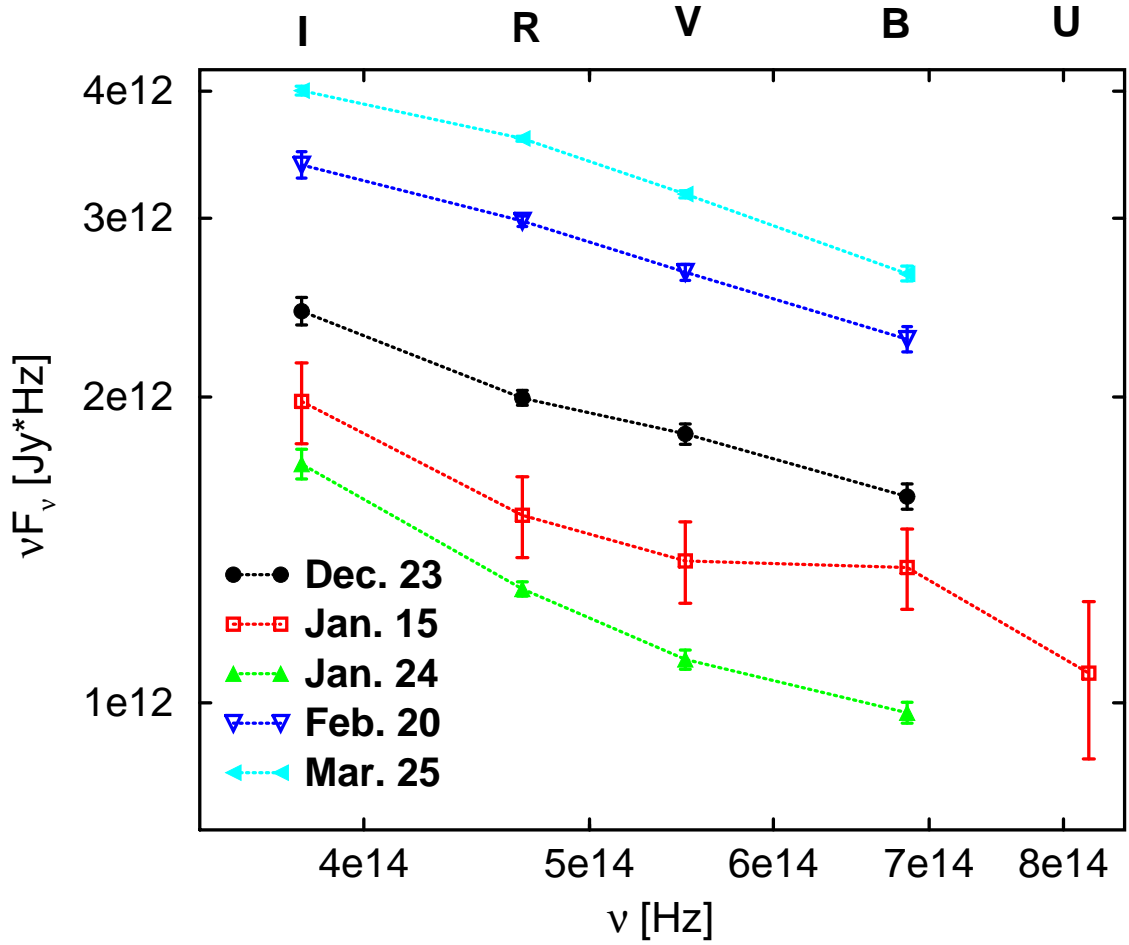}
\caption{Snap-shot optical (BVRI) continuum spectra for 5 epochs 
during our campaign. All measurements for each individual spectrum
were taken within $\le 20$~min. of each other. }
\label{opt_sp}
\end{figure}

A similar trend was recently observed in a multiwavelength WEBT campaign
on the quasar 3C~454.3 \citep{villata06}, where it could be interpreted as 
a ``little blue bump'' due to the unresolved contribution from optical emission
lines in the $\sim 2000$ -- 4000~\AA\ wavelength range in the rest frame
of the quasar, in particular from Fe~II and Mg~II \citep{raiteri07}.
In order to test whether such an interpretation would also be viable in 
the case of 3C~279, we have compiled several simultaneous snap-shot optical 
continuum (BVRI) spectra at various brightness levels (Fig. \ref{opt_sp}).
All measurements for each individual spectrum displayed in Fig. \ref{opt_sp}
have been taken within $\le 20$~min. of each other. Only the spectrum of
Jan. 15 shows a significant deviation from a pure power-law in the B
band; and in that case, there was simultaneous U band coverage, which
matched a straight power-law extrapolation of the VRIJHK spectrum. 
If the spectral upturn towards the blue end of the spectrum were 
due to an unresolved Mg~II / Fe~II line contribution, it should
emerge even more clearly in the Jan. 24 spectrum, which is characterized
by a lower optical continuum flux level than the Jan. 15 spectrum. 
Therefore, the compilation of spectra in Fig. \ref{opt_sp} does not 
provide any support for the existence of an essentially non-variable 
continuum component at the blue end of the spectrum. For this reason, 
we believe that the color changes that we found in our data are in fact 
intrinsic to the blazar jet emission.

\begin{figure}
\plotone{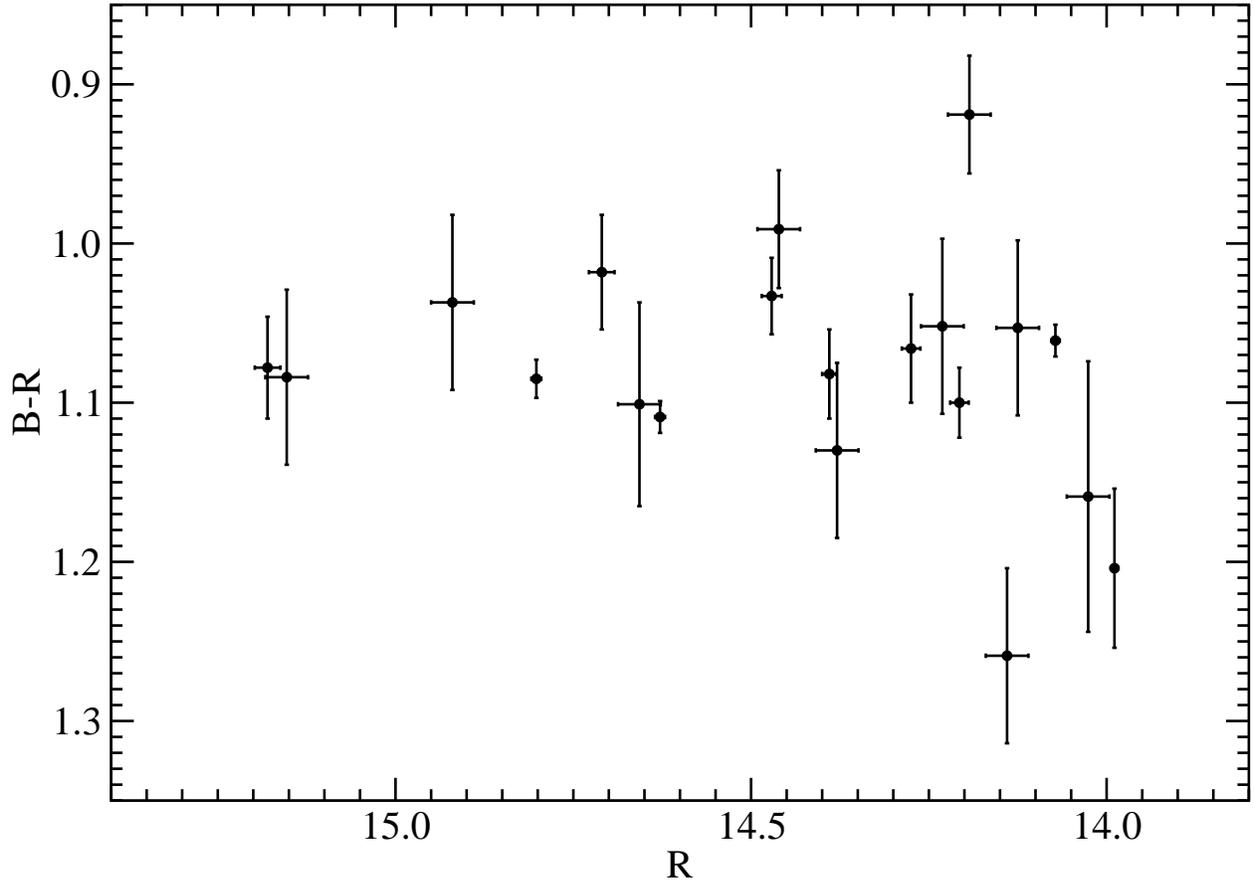}
\caption{Color-magnitude diagram for the entire campaign period.}
\label{color_mag}
\end{figure}

Fig. \ref{color_mag} illustrates our impression from Fig. \ref{B_R_lc},
that there is no clear overall trend of source (R-band) brightness with
optical spectral hardness. However, the figure clearly illustrates that
the scatter of the B -- R color index is significantly larger at larger
source brightness, indicating that spectral variability is more likely 
to occur when the source is bright. Specifically, for $R > 14.5$, the
data is consistent with a roughly constant value of B -- R = 1.09,
corresponding to a spectral index of $\alpha_o = 1.9$ for a power-law
continuum spectrum with $F_{\nu, o} [{\rm Jy}] \propto \nu^{-\alpha_o}$. 
At brightness levels $R < 14.5$, spectral variability by $\Delta (B - R)
\lesssim 0.35$, corresponding to $\Delta\alpha_o \lesssim 0.85$, is
observed.

\begin{figure}
\plotone{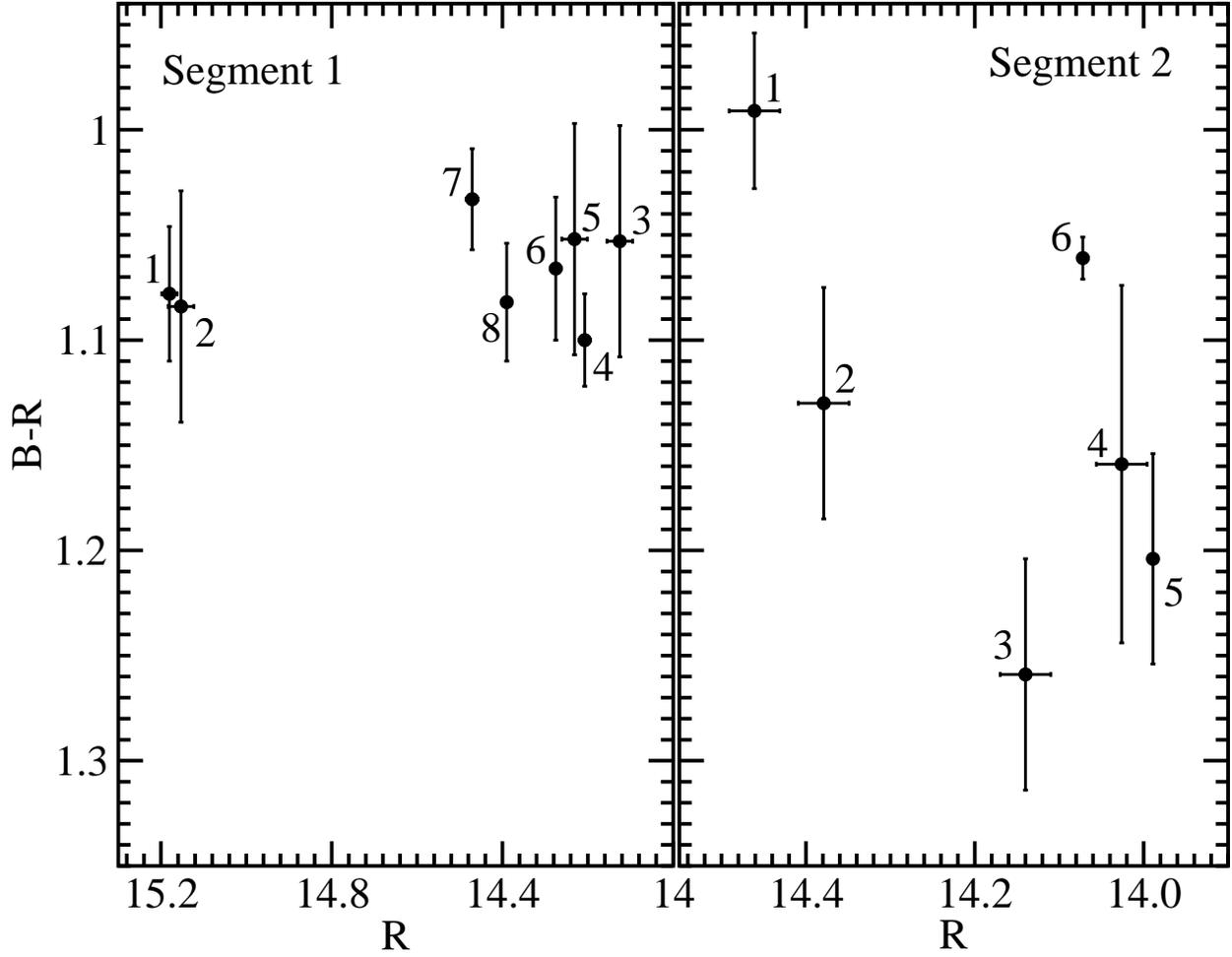}
\caption{Color-magnitude diagrams for the two time segments marked in
Fig. \ref{B_R_lc}, with time ordering indicated by the numbers in the
two panels. }
\label{hysteresis}
\end{figure}

In Fig. \ref{hysteresis}, we focus on two $\sim 1$-month optical flares,
around mid-Jan. -- mid-Feb. 2006, and late Feb. -- late March 2006,
as indicated by the gray shaded segments 1 and 2, respectively, in 
Fig. \ref{B_R_lc}. The number labels in the color-magnitude diagrams
in Fig. \ref{hysteresis} indicate the time ordering of the points.
While there is no obvious trend discernible in Segment 1, Segment 2
suggests the presence of a spectral hysteresis pattern: The spectral
softening around JD 2453790 -- 2453806 (Feb. 23 -- Mar. 11), already
mentioned above, precedes the main brightness increase. Subsequently,
the optical continuum hardens while the source is still in a bright
state. We need to caution that due to the poor sampling of the B band
light curve during segment 2, the significance of the tentative hysteresis 
found here may be questionable. However, such hysteresis would naturally
lead to a B-band time lag behind the R-band, for which we do find a
$3.9 \, \sigma$ evidence from a discrete correlation function analysis
as described in the following section.

To our knowledge, such a spectral hysteresis has never been observed
at optical wavelengths for any flat-spectrum radio quasar. It is 
reminiscent of the spectral hysteresis occasionally seen at X-ray 
energies in high-frequency peaked BL~Lac objects \citep[HBLs, 
e.g.][]{takahashi96,fossati00,kataoka00}. However, the spectral
hysteresis observed in the X-rays of HBLs is generally clockwise
(i.e., spectral hardening precedes flux rise; softening precedes
flux decline), and can be interpreted as the synchrotron signature 
of fast acceleration of ultrarelativistic electrons, followed by
a gradual decline on the radiative cooling time scale 
\citep[e.g.,][]{kataoka00,kusunose00,lk00,bc02}. In our 
case, the direction of the spectral hysteresis is counterclockwise
(i.e., spectral softening precedes the flux rise; spectral hardening
precedes flux decline). Possible physical implications of such 
hysteresis phenomena will be discussed in \S \ref{discussion}.

\begin{figure}
\plotone{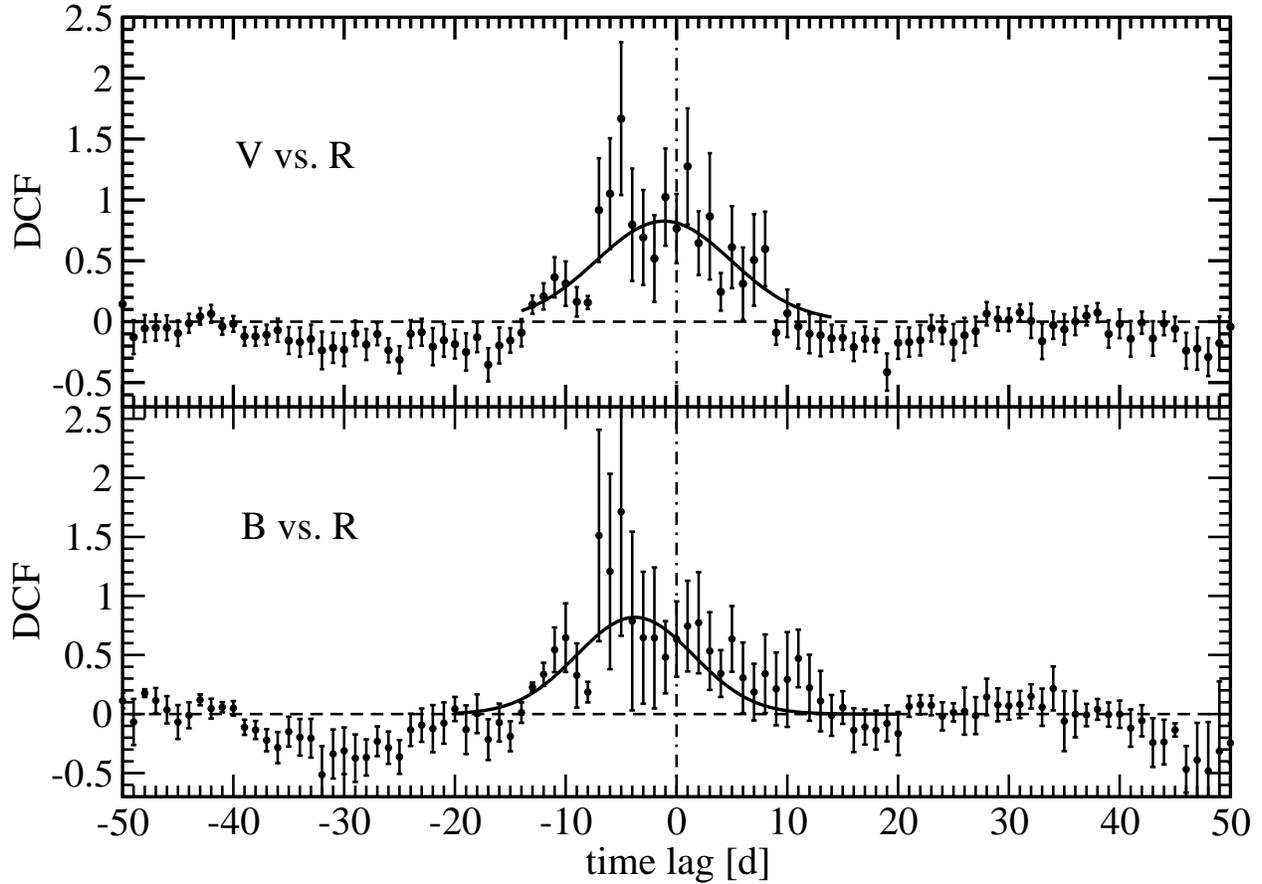}
\caption{Discrete correlation functions between V and R (top panel),
and B and R (bottom panel), respectively. The solid curves indicate
the best fit with a symmetric Gaussian. This leads to a best-fit maximum 
correlation at $\tau_0 = (-1.14 \pm 0.48)$~d for the V band and $\tau_0
= (-3.75 \pm 0.96)$~d for the B band, indicating a lag of the V and B
band light curves behind the R-band. }
\label{DCF}
\end{figure}

\section{\label{crosscorrelations}Inter-band cross-correlations and time lags}

The result of an occasional counterclockwise hysteresis in 3C~279,
as found in the previous section, immediately suggests the existence 
of a characteristic time lag of higher-frequency behind lower-frequency 
variability. In order to investigate this, we evaluated the discrete
correlation function \citep[DCF,][]{ek88} between the R band and the 
other optical light curves. In our notation, a positive value of the
time lag $\Delta \tau$ would indicate a lag of the R-band light curve 
behind the comparison light curve. As mentioned earlier, the radio (and 
near-IR) light curves are too sparsely sampled and yielded no significant 
features in the DCF. Fig. \ref{DCF} shows the DCF between the R band and 
the V band (top panel) and the B band (bottom panel), using a sampling time
scale of $\Delta\tau = 1$~d. We have done the same analysis using various
other values of $\Delta\tau$, which yielded results consistent with the
ones described below. We chose to show the results for $\Delta\tau = 1$~d
because they provided the best compromise between dense time scale sampling 
and reduction of error bars. 

The DCFs reveal clear correlations between the different optical wavebands, 
with peak values around 1. This confirms our previous impression from 
inspection by eye, that the variability patterns in all optical wavebands 
track each other very closely. 

\begin{figure}
\plotone{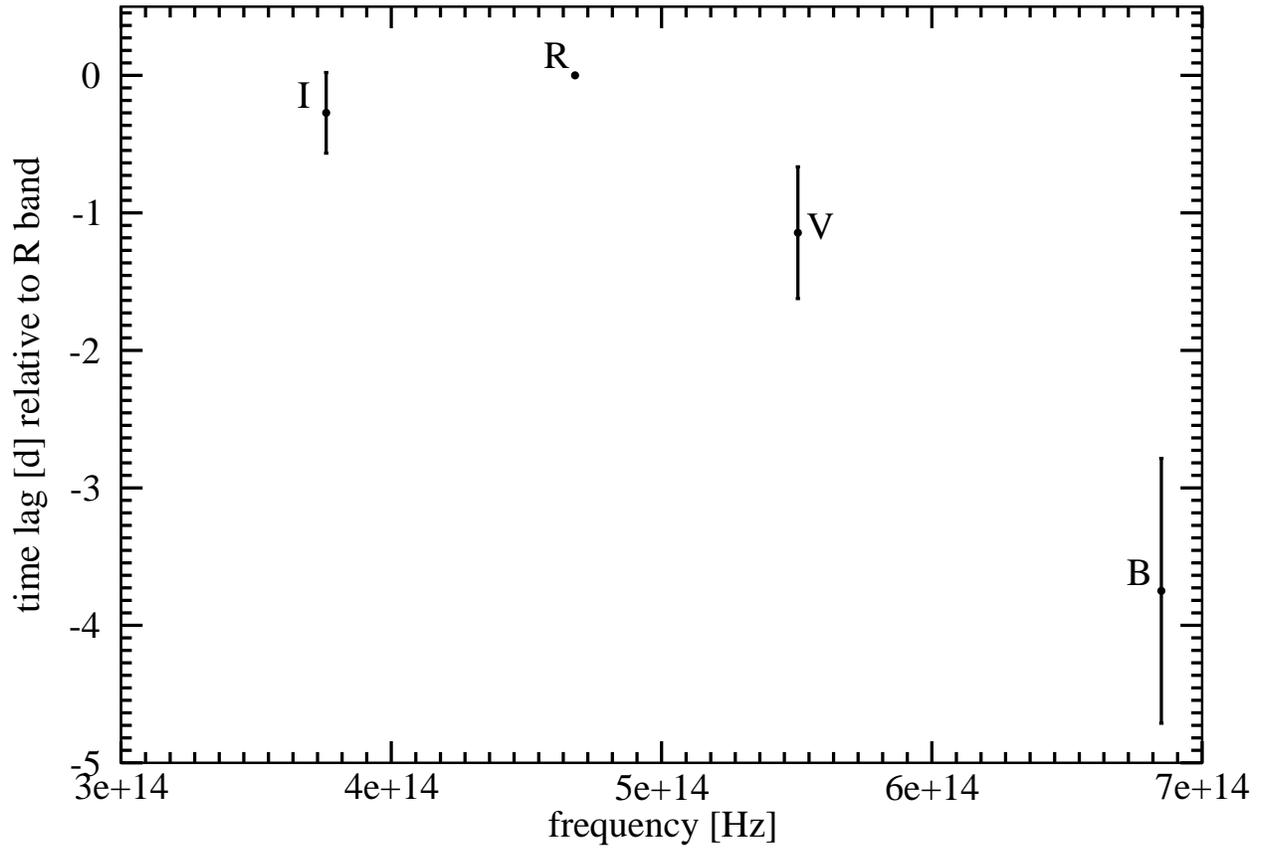}
\caption{Best-fit time lag of the R band vs. B, V, R, and I bands, as a 
function of frequency.}
\label{tau_frequency}
\end{figure}

The resulting DCFs have then been fitted with a symmetric Gaussian, 
$F_0 \, e^{-(\tau - \tau_0)^2/(2 \sigma^2)}$. This analysis yields
non-zero offsets of the best-fit maxima, $\tau_0$ at the $\sim 3$ -- 
$4 \sigma$ level. Specifically, we find $\tau_0 = (-1.14 \pm 0.48)$~d 
for the V band and $\tau_0 = (-3.75 \pm 0.96)$~d for the B band. This
indicates a hard time lag of higher-frequency variability behind the
variability at lower frequencies in the B-V-R frequency range. However,
this trend does not continue into the I band. This is illustrated in 
Fig. \ref{tau_frequency}, where we plot the best-fit time lags as a 
function of photon frequency. However, we need to add a note of caution:
The rather sparse sampling of the B- and V-band light curves leads to
large error bars on the DCFs. Clearly, alternative, more complex 
representations, e.g., multiple Gaussians and/or asymmetric functions, 
will also provide acceptable fits to the observed DCFs and may lead 
to different quantitative results concerning the involved time lags. 
Future observations with denser B- and V-band sampling are needed in 
order to test the robustness of the trend found here. 

The hard lag found in this analysis may be physically related to
the $\sim 2$ -- 3~d time lag of the soft X-ray spectral index behind 
the flux in {\it ROSAT} observations of 3C~279 in December 1992 --
January 1993 \citep{pian99}. Possible physical interpretations will
be discussed in \S \ref{discussion}.

\begin{figure}
\plotone{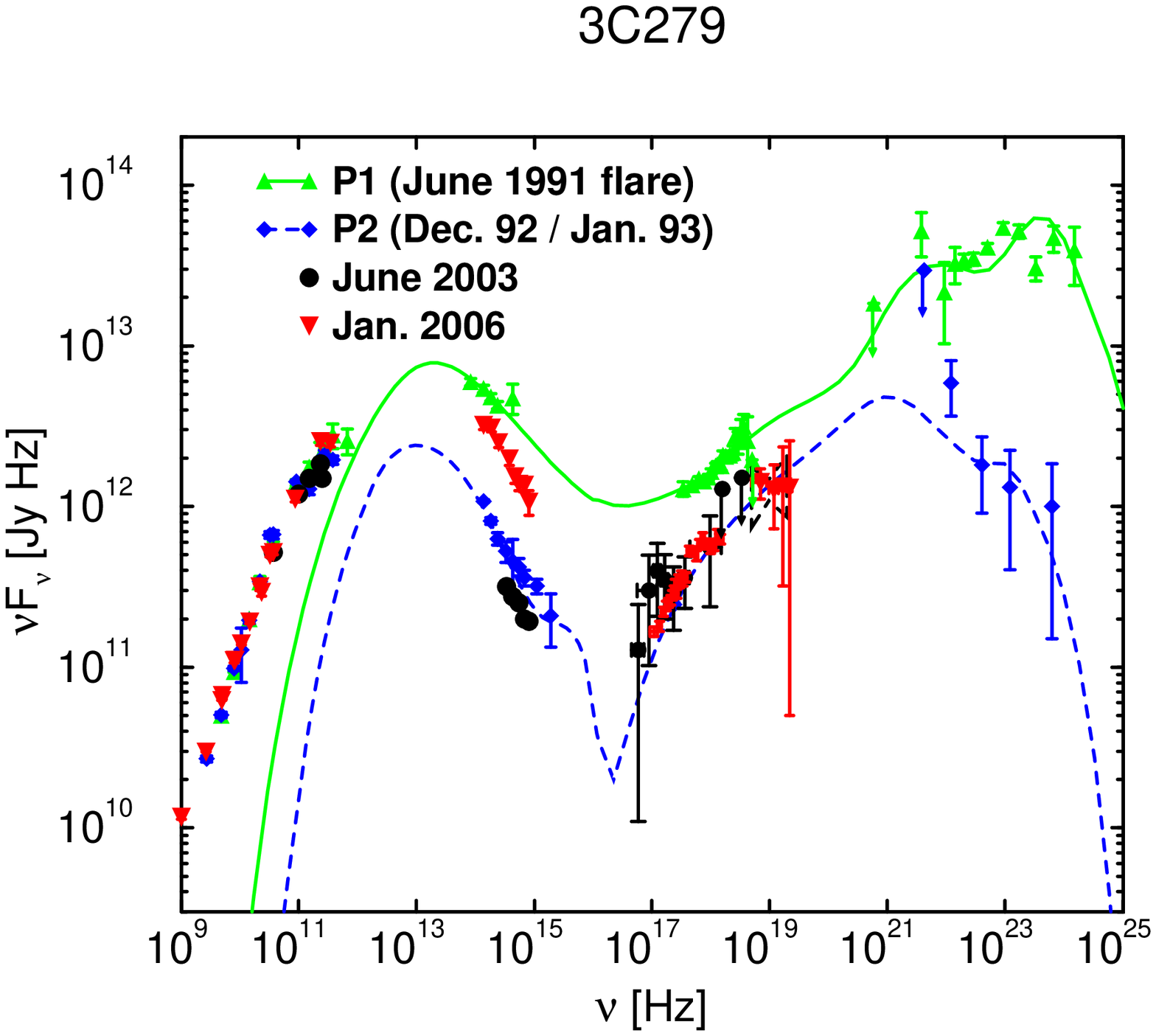}
\caption{Simultaneous snap-shot spectral energy distributions of 3C~279 at 
various epochs. Data pertaining to the multiwavelength campaign in January
2006 are plotted with red triangles. All IR and optical data were taken within 
$\pm 1/2$~hr of UT 05:00 on Jan. 15, 2006; the X-ray and soft $\gamma$-ray 
data represent the time-averaged spectra throughout the respective observing
windows in Jan. 2006, which included Jan. 15 for all instruments. Model fits
to the {\it EGRET} P1 and P2 SEDs are calculated using a time-dependent leptonic 
(SSC + EC) jet model, and taken from \cite{hartman01a}. The June 2003 SED
is from \cite{collmar04}. }
\label{SED}
\end{figure}

\section{\label{spectra}Broad-band spectral energy distribution}

The most complete broadband coverage of 3C~279 during our campaign was
obtained on January 15, 2006. On that day, the only near-infrared (JHK)
exposures were taken with the NOT, and simultaneous radio observations
at 5, 8, 22, and 37~GHz were available as well. This is also within the
time window of the X-ray and soft $\gamma$-ray observations, although
in most cases, a meaningful extraction of spectral information required 
the integration over most of the high-energy observing period, January
13 -- 20, 2006 \citep{collmar07a}. The total snap-shot SED composed of
all available radio, near-IR, optical, X-ray, and soft $\gamma$-ray 
observations around January 15, 2006 is displayed in Fig. \ref{SED}, 
where all near-IR and optical data are taken within $\pm 1/2$~hr of 
UT 05:00 on January 15. The figure compares the January 2006 SED to 
previous SEDs from the bright flare during the first {\it EGRET} 
observing epoch in June 1991, the low state of December 1992 / January 1993
\citep[both SEDs adapted from][]{hartman01a}, and a previous multiwavelength 
campaign around {\it INTEGRAL} AO-1 observations of 3C~279 in a low state
in June 2003 \citep{collmar04}. 

While the optical (presumably synchrotron) emission component clearly
indicates that the source was in an elevated state compared to previous
low states, the simultaneous X-ray -- soft-$\gamma$-ray spectrum 
is perfectly consistent with the low-state spectra of 1992/1993 and 
2003. This is a very remarkable result and will be discussed in detail
in a companion paper about the results of the entire multiwavelength
campaign \citep{collmar07b}.

The NIR -- optical continuum can be very well represented by a single
power-law with a spectral index of $\alpha_o = 1.64 \pm 0.04$, corresponding 
to an underlying non-thermal electron distribution with a spectral index
of $p = 4.28 \pm 0.08$, if the optical continuum is synchrotron emission. As
already pointed out in \cite{hartman01a} for a comparison between various
{\it EGRET} observing epochs over $\sim 10$~years and confirmed by our 
campaign data on time scales of weeks -- months (see \S \ref{variability}), 
the optical spectral index in 3C~279 does not show any systematic 
correlation with the brightness state of the source, and Fig. \ref{SED} 
indicates that in our high-state observations of 2006, the continuum 
spectral slope is not significantly different from the slopes observed 
in the low states of 1992/1993 and 2003.

\section{\label{discussion}Discussion}

In this section, we discuss some general physical implications and 
constraints that our results can place on source parameters. In the
following discussion, we will parameterize the magnetic field in units
of Gauss, i.e., $B = 1 \, B_G$~Gauss, and the Doppler boosting factor
$D = \left(\Gamma [ 1 - \beta_{\Gamma} \cos\theta_{\rm obs}] \right)^{-1}$
in units of 10, i.e., $D = 10 \, D_1$ where $\Gamma$ is the bulk Lorentz 
factor of the emitting region, $\beta_{\Gamma}$~c is the corresponding 
speed, and $\theta_{\rm obs}$ is the observing angle. The characteristic 
variability time scale is of the order of 1 -- a few days, so we write 
$t_{\rm var}^{\rm obs} \equiv 1 \, t_{\rm var, d}^{\rm obs}$. In the
same sense, we parameterize the time lag of the B band behind the R band 
as $\tau_{\rm BR}^{\rm obs} \equiv 1 \, \tau_{\rm BR, d}^{\rm obs}$~d with 
$\tau_{\rm BR, d}^{\rm obs} \sim 3$. The observed variability time scale
yields an estimate of the size of the emitting region, $R \equiv 10^{15} 
R_{15}$~cm through $R \lesssim c \, D/(1 + z) \, t_{\rm var}^{\rm obs}$.
We find $R_{15} \lesssim 17 \, D_1 \, t_{\rm var, d}^{\rm obs}$. 

The (co-moving) energies of electrons emitting synchrotron radiation at 
their characteristic peak frequencies in the R and B bands are

\begin{eqnarray} 
\gamma_R &= 3.7 \times 10^3 \, \left( B_G \, D_1 \right)^{-1/2} \cr
\gamma_B &= 4.5 \times 10^3 \, \left( B_G \, D_1 \right)^{-1/2}
\label{gamma_BR}
\end{eqnarray}

The steep underlying electron spectrum with $p = 4.5$, inferred from the 
steep optical continuum, might indicate that the entire optical spectrum
is produced by electrons in the fast-cooling regime. This implies that
the radiative cooling time scale of electrons emitting synchrotron radiation
in the optical regime is shorter than the characteristic escape time scale
of those electrons. The respective time scales in the co-moving frame,
$t'_{\rm esc}$ and $t'_{\rm cool}$ can be written as

\begin{eqnarray} 
t'_{\rm esc} &\equiv \eta \, {R \over c} & \lesssim 5.7 \times 10^5 \, \eta \,
D_1 \, t_{\rm var, d}^{\rm obs} \; {\rm s} \cr
t'_{\rm cool} &= {\gamma \over {\dot\gamma}_{\rm rad}} &\sim 7.7 \times 10^7
\, B_G^{-1} \, \gamma^{-1} \, (1 + k)^{-1} \; {\rm s}
\label{timescales}
\end{eqnarray}
where $\eta \ge 1$ is the escape time scale parameter (as defined in 
the first line of eq. \ref{timescales}), and $k$ is a correction
factor accounting for radiative cooling via Compton losses in the Thomson
regime in a radiation field with an energy density ${u'}_{\rm rad} \equiv
k \, {u'}_{\rm sy}$. Requiring that the cooling time scale is shorter than
the escape time scale, at least for electrons emitting in the R band,
leads to a lower limit on the magnetic field:

\begin{equation}
B \gtrsim B_{\rm c} \equiv 1.3 \times 10^{-3} \, (1 + k)^{-2} \, \eta^{-1} 
\, (t_{\rm var, d}^{\rm obs})^{-2} \; {\rm G}
\label{B_cooling}
\end{equation}
which does not seem to pose a severe constraint, given the values of
$B \sim$~a few G typically found for other FSRQs and also for 3C~279
from previous SED modeling analyses \citep[e.g.,][]{hartman01a}.

Another estimate of the co-moving magnetic field can be found by 
assuming that the dominant portion of the time-averaged synchrotron 
spectrum is emitted by a power-law spectrum of electrons with 
$N_e (\gamma) = n_0 \, V_B \, \gamma^{-p}$ for $\gamma_1 \le \gamma 
\le \gamma_2$; here, $V_B$ is the co-moving blob volume, and we use
$p = 4.5$ as a representative value inferred from the optical continuum 
slope. The normalization constant $n_0 = n_e \, (1 - p) / 
\left(\gamma_2^{1 - p} - \gamma_1^{1 - p}\right)$ is related to the 
magnetic field through an equipartition parameter $e_B \equiv {u'}_B / 
{u'}_e$ (in the co-moving frame). Note that this equipartition parameter 
only refers to the energy density of the electrons, not accounting for 
a (possibly dominant) energy content of a hadronic matter component in 
the jet. Under these assumptions, the magnetic field can be estimated 
as described, e.g., in \cite{boettcher03}. Taking the $\nu F_{\nu}$ peak 
synchrotron flux $f_{\epsilon}^{\rm sy}$ at the dimensionless synchrotron 
peak photon energy $\epsilon_{\rm sy} \equiv E_{\rm pk, sy}/(m_e c^2) 
\approx 3 \times 10^{-7}$ as $\sim 10^{-10}$~ergs~cm$^{-2}$~s$^{-1}$, 
we find

\begin{equation}
B \gtrsim B_{\rm e_B} \equiv 1.86 \, D_1^{-13/7} \, e_B^{2/7} \, 
(t_{\rm var, d}^{\rm obs})^{-1} \; {\rm G}.
\label{B_eB}
\end{equation}

This constitutes a more useful and realistic magnetic-field estimate than 
eq. \ref{B_cooling}. If, indeed, the optical emission is synchrotron emission
from a fast-cooling electron distribution, then electrons have been primarily
accelerated to a power-law distribution with an injection index of $q = p - 1
= 3.5$. This is much steeper than the canonical spectral index of $q \sim 2.2$
-- 2.3 found for acceleration on relativistic, parallel shocks 
\citep[e.g.,][]{gallant99,achterberg01}, and could indicate an oblique
magnetic-field orientation \citep[e.g.,][]{ob02,no04}, which would yield
a consistent picture with the predominantly perpendicular magnetic-field
orientation observed on parsec-scales \citep{jorstad04,ojha04,lh05}. The
observed hard lag (B vs. R) may then be the consequence of a gradual spectral 
hardening of the electron acceleration (injection) spectrum throughout the
propagation of a relativistic shock front along the jet. Such a gradual 
hardening of the electron acceleration spectrum could be the consequence
of the gradual build-up of hydromagnetic turbulence through the relativistic
two-stream instability \citep[see, e.g.,][]{schlickeiser02}. This
turbulence would harden the relativistic electron distribution via 
2nd-order Fermi acceleration processes \citep{vv05}. Such a scenario 
would imply a length scale for the build-up of turbulence of
$\Delta r \sim c \, \tau_{\rm BR}^{\rm obs} \, D \, \Gamma / (1 + z)
\sim 5.6 \times 10^{-2} \, \tau_{\rm BR, d}^{\rm obs} \, D_1 \, \Gamma_1
\sim 0.2$~pc for the characteristic values of 3C~279. 

Alternatively, the acceleration process could become more efficient along
the jet if the magnetic-field configuration gradually evolves into a 
more quasi-parallel one, on the same length scale of $\sim 0.2$~pc as
estimated above. However, this scenario might be in conflict with the 
predominantly perpendicular magnetic-field orientation seen in the jets 
of 3C~279 on pc scales.

The hard lag in the optical regime may also be indicative of a slow
acceleration mechanism, with an acceleration time scale of the order
of the observed B vs. R lag. This would imply an acceleration rate of

\begin{equation}
\dot\gamma_A \sim {\gamma_B - \gamma_R \over \tau_{\rm BR}}
\sim 6.8 \times 10^{-2} (\tau_{\rm BR, d}^{\rm obs})^{-1} \,
\left( {D_1 \over B_G} \right)^{1/2} \; {\rm s}^{-1}.
\label{acceleration}
\end{equation}
In this scenario, electrons could only be accelerated to at least $\gamma_B$,
if the acceleration rate of eq. \ref{acceleration} is larger than the absolute 
value of the radiative (synchrotron + Compton) cooling rate corresponding to eq.
\ref{timescales}. This imposes an upper limit on the magnetic field:

\begin{equation}
B \lesssim B_{\rm acc} \sim 0.42 \, D_1^{1/3} \, (\tau_{\rm BR, d}^{\rm obs})^{-2/3}
\, (1 + k)^{-2/3} \; {\rm G}.
\label{B_acc}
\end{equation}
This can be combined with the estimate in eq. \ref{B_eB} to infer a limit
on the magnetic-field equipartition parameter:

\begin{equation}
e_B \lesssim e_{B, acc} \sim 5.8 \times 10^{-3} \, D_1^{23/3} \, 
(\tau_{\rm BR, d}^{\rm obs})^{-7/3} \, (t_{\rm var, d}^{\rm obs})^{7/2} \,
(1 + k)^{-7/3}.
\label{eB_acc}
\end{equation}
Based on this equipartition parameter, one can use the magnetic-field
estimate of eq. \ref{B_eB} to estimate the total amount of co-moving 
energy contained in the emission region at any given time:

\begin{equation}
E'_e \sim {4 \over 3} \, \pi \, R^3 \, {{u'}_B \over e_B}
\sim 2.5 \times 10^{49} \, D_1^{-4} \, \tau_{\rm BR, d}^{\rm obs}
\, (t_{\rm var, d}^{\rm obs})^{-1/2} \, (1 + k) \; {\rm erg}
\label{energy}
\end{equation}
Assuming that the bulk of this energy is dissipated within the 
characteristic variability time scale, one can estimate the power
in relativistic electrons in the jet:

\begin{equation}
L_{\rm jet} \sim {E'_e \over {t'}_{\rm var}} \sim 4.5 \times 10^{43}
\, D_1^{-5} \, \tau_{\rm BR, d}^{\rm obs} \, (t_{\rm var, d}^{\rm obs})^{-3/2} 
\, (1 + k) \; {\rm ergs \; s}^{-1}.
\label{Ljet}
\end{equation}

Previous modeling works of the SEDs of FSRQs in general and 3C~279 in
particular indicated characteristic magnetic field values of a few Gauss,
in approximate equipartition with the ultrarelativistic electron population.
The unusually low equipartition parameter in eq. \ref{eB_acc} could therefore
pose a problem for the slow-acceleration scenario. Note, however, the very
strong dependence of $e_B$ on the Doppler factor ($\propto D^{23/3}$). A
Doppler factor $D \sim 20$ could account for equipartition parameters of 
the order of one. Also, the energy requirements of eqs. \ref{energy} and 
\ref{Ljet} seem reasonable, and there does not appear to be a strict argument 
that would rule this scenario out. 

Another scenario one could think of would be based on a decreasing magnetic 
field along the blazar jet, leading to a gradually increasing cooling
break in the underlying electron distribution. This would require that the
cooling time scale for electrons emitting synchrotron radiation in the
optical regime would be equal to or longer than the escape time scale.
Thus, the inequality in eq. \ref{B_cooling} would be reversed. This would 
require unreasonably low magnetic fields. Furthermore, this scenario would 
be in conflict with the typically observed unbroken snap-shot power-law 
continuum spectra throughout the optical-IR range. Therefore, this idea
may be ruled out.

\section{\label{summary}Summary}

We have presented the results of an optical-IR-radio monitoring campaign 
on the prominent blazar-type flat-spectrum radio quasar 3C~279 by the 
WEBT collaboration in January -- April 2006, around Target of Opportunity 
X-ray and soft $\gamma$-ray observations with {\it Chandra} and {\it INTEGRAL} 
in mid-January 2006. Previously unpublished radio and optical data from 
several weeks leading up to the ToO trigger are also included. 

The source exhibited substantial variability of flux and spectral shape, 
in particular in the optical regime, with a characteristic time scale
of a few days. The variability patterns throughout the optical BVRI 
bands were very closely correlated with each other, while there was 
no significant evidence for a correlation between the optical and 
radio variability. After the trigger flux level for the {\it Chandra} 
and {\it INTEGRAL} ToOs was reached on Jan. 5, 2006, the optical 
flux decayed smoothly by 1.1 mags. within 13 days, until the end 
of the time frame of the X-ray and $\gamma$-ray observations. The 
decay could be well described by an exponential decay with a 
decay time scale of $\tau_d = 12.8$~d. The flux then recovered 
to approximately the pre-dip values in a much more erratic way, 
including a $\sim 0.5^{\rm mag}$ rise within $\sim 1$~d. 

A discrete correlation function analysis between different optical 
(BVRI) bands indicates a hard lag with a time delay increasing with 
increasing frequency, reaching $\sim 3$~d for the lag of B behind 
R. This appears to be accompanied by a single indication of 
counterclockwise spectral hysteresis in a color-intensity diagram 
(B-R vs. R). Thus, spectral hardening during flares appears delayed 
with respect to a rising optical flux. There is no consistent
overall trend of optical spectral hardness with source brightness.
However, our data indicate that the source displays a rather uniform
spectral slope of $\alpha_o \sim 1.9$ at moderate flux levels ($R > 
14.5$), while spectral variability seems common at high flux levels
($R < 14.5$). 

The occasional optical spectral hysteresis, in combination with the 
very steep IR-optical continuum spectral index of $\alpha_o \sim 1.5$
-- 2.0, may indicate a highly oblique magnetic field configuration near 
the base of the jet, leading to inefficient particle acceleration and a 
very steep electron injection spectrum. As the emission region propagates 
along the jet, a gradual hardening of the primarily injected ultrarelativistic 
electron distribution may be caused by the gradual build-up of hydromagnetic 
turbulence, which could lead to a gradually increasing contribution of 
second-order Fermi acceleration. This would imply a length scale of the
build-up of hydromagnetic turbulence of $\Delta r \sim 0.2$~pc. 

An alternative explanation of the hard lag may be a slow acceleration 
mechanism by which relativistic electrons are accelerated on a time scale 
of several days. However, even though this model can plausibly explain the 
observed variability trends and overall luminosity of the source, it requires 
an unusually low magnetic field in the emitting region of $B \lesssim 0.2$~G,
unless rather high Doppler factors of $D ~ 20$ are assumed. Such a small 
magnetic field would be about an order of magnitude lower than inferred 
from previous analyses of simultaneous SEDs of 3C~279 and other flat-spectrum 
radio quasars with similar properties.

\acknowledgments
The work of M. B\"ottcher and S. Basu was partially supported by 
NASA through INTEGRAL GO grant award NNG~06GD57G and the Chandra
GO program (administered by the Smithsonian Astrophysical Observatory)
through award no. GO6-7101A.
The Mets\"ahovi team acknowledges the support from the Academy of
Finland.
YYK is a research fellow of the Alexamder von Humboldt Foundation.
RATAN-600 observations were partly supported by the Russian Foundation
for Basic Research (project 05-02-17377).
The St. Petersburg team was supported by the Russian Foundation for
Basic Research through grant 05-02-17562.

\end{document}